\documentclass[12pt,preprint]{aastex}
\usepackage[numberedappendix]{emulateapj5}  

\slugcomment{submitted to ApJ} 
 
\shorttitle{AGN heating in clusters}
\shortauthors{Br\"ggen et al.}
 
\received{2004 December}
\begin{document}

%
%
\def\msun{${\rm M_{\odot}} \;$}
\def\be{\begin{equation}}
\def\ee{\end{equation}}
\def\gcc{gcm$^{-3}$}
\def\bi{\begin{itemize}}
\def\ei{\end{itemize}}
\def\Mo{$M_{\odot}$}
\def\Lo{$L_{\odot}$}
\def\s{s$^{-1}$}  
 
\title{AGN heating and dissipative processes in galaxy clusters}
\author{M. Br\"uggen}
\affil{International University Bremen, Campus Ring 1, 28759 Bremen,
Germany; \email{m.brueggen@iu-bremen.de}}
\author{M. Ruszkowski\altaffilmark{1}}
\affil{Joint Institute for Laboratory Astrophysics, Campus Box 440, University of Colorado at Boulder, CO
80309-0440; \email{mr@quixote.colorado.edu}}               
\altaffiltext{1}{{\it Chandra} Fellow}
\and
\author{E. Hallman}
\affil{Center for Astrophysics and Space Astronomy, University of Colorado 
at Boulder, CO 80309; \email{hallman@origins.colorado.edu} }

\begin{abstract}
  Recent X-ray observations reveal growing evidence for heating by active
  galactic nuclei (AGN) in clusters and groups of galaxies.  AGN outflows play
  a crucial role in explaining the riddle of cooling flows and the entropy
  problem in clusters.  Here we study the effect of AGN on the intra-cluster
  medium in a cosmological simulation using the adaptive mesh refinement FLASH
  code.  We pay particular attention to the effects of conductivity and
  viscosity on the dissipation of weak shocks generated by the AGN activity in
  a realistic galaxy cluster.  Our 3D simulations demonstrate that both
  viscous and conductive dissipation play an important role in distributing
  the mechanical energy injected by the AGN, offsetting radiative cooling and
  injecting entropy to the gas. These processes are important even when the
  transport coefficients are at a level of 10\% of the Spitzer value. Provided
  that both conductivity and viscosity are suppressed by a comparable amount,
  conductive dissipation is likely to dominate over viscous dissipation.
  Nevertheless, viscous effects may still affect the dynamics of the gas and
  contribute a significant amount of dissipation compared to radiative
  cooling. We also present synthetic {\it Chandra} observations.  We show that
  the simulated buoyant bubbles inflated by the AGN, and weak shocks
  associated with them, are detectable with the {\it Chandra} observatory.
\end{abstract}


\section{Introduction}
Clusters of galaxies are excellent laboratories for studying the
  interaction between outflows from active galactic nuclei with
  diffuse baryons.  Recent observational evidence demonstrates that
  the lives of AGN and galaxy clusters in which they reside are
  closely intertwined.  In particular, there is mounting observational
  evidence that AGN may provide a vital clue in explaining the cooling
  flow problem. In the absence of non-gravitational heating, the
  cluster cores should cool and accrete gas at rates of hundreds and
  more solar masses per year.  However, this is in conflict with
  observational evidence which often indicates mass deposition rates
  of only $\sim 10$ solar masses per year. Moreover, the gas
  temperatures in cluster centers are typically maintained above $\sim
  2$ keV.  As many cooling flow clusters are known to harbor active
  radio sources (Burns 1990, Eilek 2004) and the enthalpy of cavities
  inflated by radio galaxies scales with the cooling flow luminosity
  (B\^irzan et al. 2004), AGN may serve as a viable heating source to
  prevent the gas from cooling and accreting at excessive
  rates. Direct evidence for AGN heating has indeed been found in
  recent studies.  Observations of the Perseus cluster (Fabian et
  al. 2003a,b) and the Virgo cluster (Forman et al. 2004) reveal sound
  waves and weak shocks in the intra-cluster medium (ICM).  Even more
  recently, Nulsen et al. (2004) found evidence for shock heating in
  Hydra A, McNamara et al. (2004) in MS0735.6+7421 and Sanderson,
  Finoguenov \& Mohr (2004) in Abell 478. These results strongly
  suggest that AGN outflows can heat the ICM in a spatially
  distributed fashion, which may help to maintain ICM stability
  against radiative cooling. \\

\indent The evidence for non-gravitational heating in clusters has also been
observed in a statistical sense in cluster scaling relations. These relations
show departures from the self-similar scalings: In the absence of
  non-gravitational heating and radiative cooling, one would expect the
  entropy to scale proportionally with the mean cluster temperature. However,
  observations by Ponman, Sanderson \& Finoguenov (2003), Pratt \& Arnaud
  (2005) and Piffaretti et al. (2005) indicate a scaling of entropy roughly
  according to $T^{2/3}$. Moreover, they reveal a systematic excess of entropy
  at large radii in low-mass clusters (e.g., Ponman, Sanderson \& Finoguenov
  2003). More observational evidence for cluster heating comes from a number
  of massive clusters that have been shown (McCarthy et al. 2004) to host very
  large cores which require an enormous amount of heating in order to explain
  their properties. In a recent study, Croston et al. (2004) separated a
sample of groups into radio quiet and radio loud objects. They demonstrated
that radio loud groups deviate from self-similatity. This result demonstrates
that AGN play a crucial role in heating the ICM and may offer the solution to
the
entropy excess problem.\\

\indent Based on observations of the Perseus cluster it first has been
suggested by Fabian et al. (2003a,b) that viscosity may play an important role
in dissipating energy injected by the central AGN.  This idea has been tested
in numerical simulations by Ruszkowski et al. (2004a,b) and Reynolds et al.
(2005). Apart from viscosity, thermal conduction has been considered by a
number of authors as means of transferring energy from hot outer layers of
clusters towards the cool cluster cores (e.g., Fabian, Voigt, \& Morris 2002,
Voigt \& Fabian 2004, Narayan \& Medvedev 2001, Br\"uggen 2003).\\

\indent The above observational results motivate our efforts to study the
interaction between AGN and the ICM in detail.  Numerical simulations of hot,
underdense bubbles in clusters of galaxies have been performed by a number of
authors (e.g. Churazov et al. 2001, Quilis et al. 2001, Br\"uggen et al.
2002a,b, Reynolds, Heinz \& Begelman 2002, Ruszkowski, Br\"uggen \& Begelman
2004a,b, Dalla Vecchia et al. 2004). Common to these simulations is that they
use a spherically symmetric, analytical profile for the ICM. Here, in order to
get a more realistic representation of the dissipative motions in the ICM, we
perform hydrodynamical simulations of AGN heating in a cluster that has been
extracted from a cosmological simulation. Our simulations include a
  dynamic dark matter that is represented by collisionless particles and more
sophisticated gas physics.  The primary objectives of the current paper are
to study the effect of viscosity and conduction on {\it dissipation} of
AGN-induced gas motions and to assess the detectability of the signatures of
AGN heating in synthetic {\it Chandra} observations.\\

\indent The paper is organized as follows. In Section 2 we discuss the initial
conditions and give the basic informations about the setup of the simulations.
In Section 3, we discuss the implementation of heating by conduction and
viscosity. Section 4 presents the method used to simulate {\it Chandra}
observations. The results are discussed in Section 5 and summary in Section 6.

\section{Initial conditions and the simulation setup}

\begin{figure*}
\begin{center}
\includegraphics[width=6.0in,angle=0]{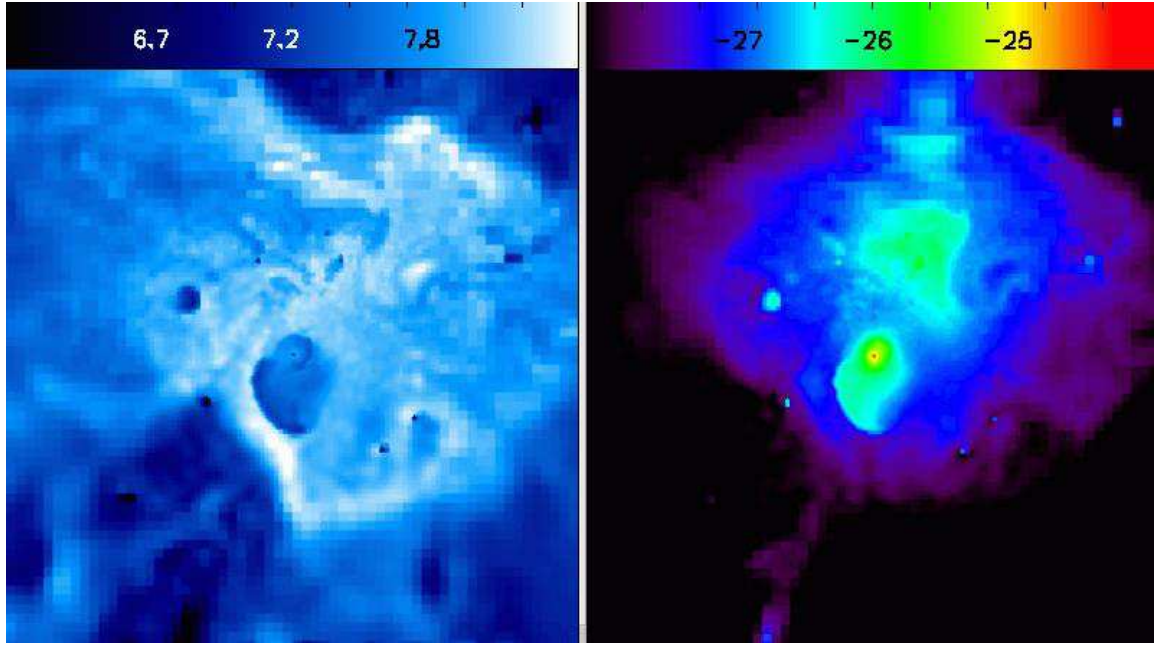}
\caption{The initial temperature (left panel) and density distribution
(right panel). Both panels show slices through the center of the
cluster and are 2.8 Mpc on a side. Plots are logarithmic and colorbars
indicate the logarithm in cgs units.}
\label{fig:initT}
\end{center}
\end{figure*}

\begin{figure*}
\begin{center}
\includegraphics[width=6.0in,angle=0]{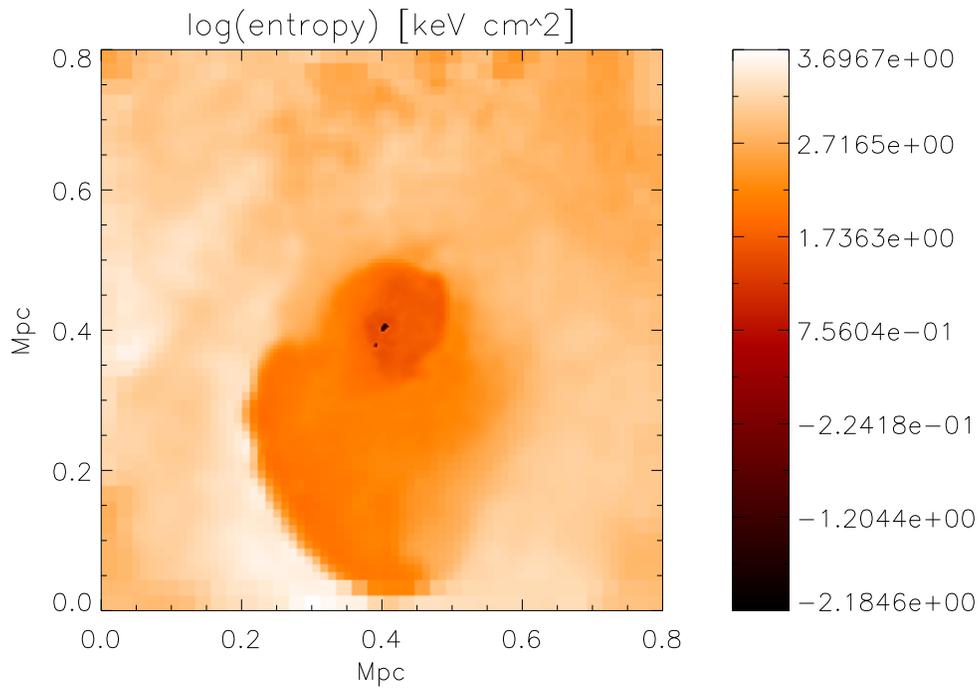}
\caption{The initial entropy distribution in a slice through the center of the
cluster. The contours are logarithmic and the colorbar
indicates the logarithm in keV cm$^2$.}
\label{fig:initentr}
\end{center}
\end{figure*}

The initial conitions for our simulations were computed with the SPH code
GADGET in standard $\Lambda$CDM cosmology ($\Omega_{\Lambda}= 0.7$,
$\Omega_{\rm m}= 0.3$, $h= 0.7$). They are based on a re-run of the S2 cluster
in Springel et al. (2001) which we chose because of all simulated
  clusters available to us its properties were closest to those of the Perseus
  cluster (in terms of temperature profile, mass and baryon fraction). At
redshift $z=0$ this cluster has a mass of $7\cdot 10^{14}$ \msun and a
  central temperature of $\sim 6$ keV.  The initial density (right panel) and
temperature (left panel) slices through the cluster center are shown in Figure
\ref{fig:initT}.  In Figure \ref{fig:initentr} we show the initial entropy
distribution. One can clearly see that this cluster is quite dynamic and shows
a lot of substructure.  It can also be seen in the temperature map that the
central part of the cluster shows a cool core characteristic of cooling flow
clusters.

The SPH simulation of the cluster includes radiative cooling and star
formation. The output of the SPH simulation serves as initial model
for our adaptive mesh refinement (AMR) simulation.  We use the FLASH
code which is a modular block-structured AMR code, parallelised using
the Message Passing Interface (MPI) library. It solves the Riemann
problem on a Cartesian grid using the Piecewise-Parabolic Method (PPM)
and, in addition, includes particles that represent the collisionless
dark matter. The particles are advanced using a cosmological
variable-timestep leapfrog-method.  Our simulations included 714346
collisionless particles that represent stars and dark matter. Gravity
is computed by solving Poisson's equation with a multigrid method
using isolated boundary conditions. For the relatively short physical
time of the bubble simulation, radiative cooling and star formation
are neglected, even though they were included in the cosmological SPH
simulation with which the cluster was produced. The central
cooling time in our cluster is about 700 Myrs which is much longer
than the 140 Myrs of our simulation. So this is probably a fair
assumption. The computational domain of our AMR simulation is a cubic
box of side $L=2 h^{-1}$ Mpc. For our grid, we chose a block size of
$16^3$ zones and used outflow boundary conditions. The minimal level
of refinement was set to 3 which means that the minimal grid contains
$16 \cdot 2^{(3-1)} = 64^3$ zones. The maximum level of refinement was
7, which corresponds to an effective grid size of $[16 \cdot
2^{(7-1)}]^{3} = [1024]^{3}$ zones or an effective resolution of $1.96
h^{-1}$ kpc. This resolution was chosen in order to capture the
relevant physical scales. Comparison with a run at a maximum
refinement level of 8 (which we could afford only for a much shorter
time) showed relatively little difference in density and temperature
and thus we think that we are close to convergence. Higher resolution
tends to yield steeper waves and thus higher dissipation.  Therefore,
one may be confident that this simulation does not overestimate the
amount of dissipation in the core. The code was run on 64 processors
of the IBM p690 at the John-von-Neumann Institut for Computing in
J\"ulich, Germany and at the National Center for Supercomputing
Applications at the Univeristy of Illinois on an identical machine.

\section{Heating}

\subsection{The energy and momentum equations}

The evolution of internal energy is followed by solving the energy equation

\begin{equation}
\rho\frac{de}{dt} = -p\Delta + \rho\dot{\epsilon}_{\rm visc} + 
                     \frac{\partial}{\partial x_{i}}\left(\kappa\frac{\partial T}{\partial x_{i}} \right),
\end{equation}

\noindent
where $\kappa = 5.0\times 10^{-7}(\ln\Lambda/37)^{-1}T^{5/2}f_c$ (Spitzer
1962) is the conductivity coefficient and $f_{c}$ is the conductivity
suppression factor.  The dissipation of mechanical energy due to viscosity,
per unit mass of the fluid, is given by (Batchelor 1967, Shu 1992, Landau \&
Lifshitz 1997)

\begin{equation}
\dot{\epsilon}_{\rm visc}=\frac{2\mu}{\rho}\left(e_{ij}e_{ij}-\frac{1}{3}\Delta^{2}\right),
\end{equation}

\noindent
where $\Delta =e_{ii}$ and

\begin{equation}
e_{ij}=\frac{1}{2}\left(\frac{\partial v_{i}}{\partial x_{j}}+\frac{\partial v_{j}}{\partial x_{i}}\right),
\end{equation}

\noindent
and where $\mu$ is the dynamical coefficient of viscosity. We use the
standard Spitzer viscosity for an unmagnetized plasma (Braginskii
1958), for which $\mu = 6.0\times
10^{-17}(\ln\Lambda/37)^{-1}T^{5/2}f_{v}$ g cm$^{-1}$ s$^{-1}$, where
$f_{v}$ is the viscosity suppression factor. We assume that both
$f_{v}$ and $f_{c}$ are equal to 0.1. It is not certain that the
suppression factors of both transport processes should be the
same. Whether they are the same may depend on the scale magnetic
fluctuations extend to. This scale may be much larger than the
gyroradii of electrons and ions (in which case suppression factors
could be comparable) or it could be comparable to the ion gyroradius
(E. Zweibel, private communication). Nevertheless, we chose the same
suppression factors for the sake of simplicity and the lack of a
better choice. The magnitude of the suppression factor is motivated by
various theoretical arguments (e.g., given in Narayan \& Medvedev
(2001) or by MHD simulations by Maron et al. 2004). However, we note
that the precise value of the suppression factor is highly uncertain
and, depending on the nature of magnetic turbulence, may even exceed
the Spitzer value (Cho et al. 2003) or be supressed well below it. The
supression factors may also vary as a function of distance from the
cluster center. An observational constraint on conductivity ($f_{c}\la
0.15$) comes from the fact that clusters would have cooled
substantially over a Hubble time if conduction had not beed supressed
(Loeb 2002).\\ 

\indent As conditions inside the buoyantly rising
bubbles are very uncertain and because we want to focus on energy
dissipation in the ambient ICM, we assume that dissipation occurs only
in regions outside the bubbles.  To this end we impose a condition
that switches on viscous and conductive effects provided that the
fraction of the injected gas in a given cell is smaller than
$10^{-1}$. \\ 

\indent Velocity diffusion was simulated by solving the
momentum equation

\begin{equation}
\frac{\partial (\rho v_{i})}{\partial t}+
\frac{\partial}{\partial x_{k}}(\rho v_{k}v_{i})+
\frac{\partial P}{\partial x_{i}} = \rho g_{i}+
\frac{\partial\pi_{ik}}{\partial x_{k}} ,
\end{equation}

\noindent
where

\begin{equation}
\pi_{ik}=\frac{\partial}{\partial x_{k}}\left[2\mu 
\left(e_{ik}-\frac{1}{3}\Delta\delta_{ik}\right)\right]
\end{equation}

\noindent
and where all other symbols have their usual meaning.

\subsection{Energy injection by AGN}
The AGN is assumed to sit in the centre of the cluster. The
intermittency period of the AGN, i.e. the period between two
subsequent bursts, was set to $3\cdot 10^7$ years, within which the
AGN was active only for a period of $5\cdot 10^5$ years. This choice
was made in order to produce a number of strong waves in quick
succession in order to gauge the dissipated energy within a feasible
simulation time. The time-averaged luminosity of the AGN was $8.3\cdot
10^{44}$ erg s$^{-1}$ and the energy was injected into two spherical
regions of radius 13 kpc that lie at a distance of 30 kpc on either
side of the centre of the cluster. This was done to simulate a bipolar
outflow that is typically seen in clusters that show signs of AGN
activity.  The energy is injected by increasing the internal energy of
the gas inside the injection region. We do not alter the velocity of
the gas inside the injection region. The injected energy is
uniformly distributed inside each injection region. The results are
insensitive to details of the spatial profile of energy injection
inside the injection regions.

\subsection{Dissipation of the injected energy}

The main transport processes responsible for the dissipation of the
injected energy that we consider in this paper are viscosity and
conductivity.  Below we discuss their relative importance in the case
of an unmagnetized plasma.\\ 

\indent Viscosity of the gas is due to off-diagonal terms in the
stress tensor in the Navier-Stokes equations and, as such, is related
to the momentum transport.  >From order of magnitude estimates we get
that viscous forces $\sim\mu \upsilon_{gas}/l^{2}$, where
$\upsilon_{gas}$ is the typical gas velocity, $l$ is the lengthscale
over which the gas properties change and $\mu$ is the coefficient of
viscosity. This coefficient is approximately given by $\mu\sim
\upsilon_{i}\lambda_{i}m_{i}n$, where $n$ is the number density of gas
particles, $\upsilon_{i}$ is a typical velocity of species $i$ (either
electrons $e$ or protons $p$) and $\lambda_{i}$ and $m_{i}$ are their
deflection length and mass, respectively. The deflection length is
independent of $m_{i}$\footnote{in the case of ionized hydrogen,
deflection time for $e^{-}-e^{-}$ collisions is shorter than for $p-p$
collisions but this is compensated by higher electron velocities}
(e.g., Frank, King and Raine 1995).  Since $\upsilon_{i}\propto
m_{i}^{-1/2}$, this implies that the coefficient of viscosity is
$\mu\propto m_{i}^{1/2}\sim m_{p}^{1/2}$ and that the transport of
momentum and viscosity are dominated by protons.\\

\indent Both electrons and protons carry the same amount of energy. As
electrons move faster than protons by $(m_{p}/m_{e})^{1/2}$, the
transfer of heat is mostly due to electrons. Thus, conductivity is
dominated by electrons and the coefficient of conductivity is
$\kappa\propto m_{e}^{-1/2}$.\\ \indent Let us now compare the
relative contribution to dissipation from the above two processes.
>From order of magnitude estimates (c.f. equations 1 and 2) we have
that $Q_{\rm cond}\sim (\kappa/k_{B})e_{\rm int}/l^{2}$ and $Q_{\rm
visc}\sim (\mu/m_{p})e_{\rm kin}/l^{2}$, where $e_{\rm int}$ and
$e_{\rm kin}$ are the internal and kinetic energies of the gas and
$k_{B}$ is the Boltzmann constant. Equipartition ensures that the
kinetic and internal energies are comparable.  Thus, $q\equiv Q_{\rm
visc}/Q_{\rm cond}\sim (k_{B}/m_{p})(\mu/\kappa)\propto
(m_{e}/m_{p})^{1/2}$.  In an idealized case of plane linear waves in a
uniform backround and without gravity the ratio of viscous to
conductive dissipation turns out to be

\begin{equation}
q=\frac{4}{3}\frac{\mu}{\kappa}\left(\frac{1}{c_{v}}-\frac{1}{c_{p}}\right)^{-1},
\end{equation}

\noindent
where $c_{v}$ and $c_{p}$ are the specific heat at constant volume and
pressure, respectively (Landau \& Lifshitz, 1997). Substituting values
appropriate for a fully ionized hydrogen plasma
$q=4.2(m_{e}/m_{p})^{1/2}\sim 0.1$ (Braginskii 1958, Spitzer 1962).
Thus, in the average sense, the effect of conductivity should dominate
over viscosity in terms of the energy dissipation.  In other words,
this says that the ratio of the P\'eclet number to the Reynolds number
is small.  In the presence of magnetic field, both, viscosity and
conductivity are thought to be suppressed. However, the Larmor radius
of ions is $(m_{i}/m_{e})^{1/2}$ times larger that the Larmor radius
of electrons of the same temperature. This can increase the P\'eclet
number relative to the Reynolds number but their ratio may to exceed
unity.\\ 

\indent We point out that, in this idealized case, the dissipation
length due to conductivity is shorter than that due to viscosity by a
factor of $q^{-1}$.  However, we also note that viscosity still plays
a role in ``diffusing'' the gas momentum by exerting viscous stress
forces and affects the overall dynamics of the gas. Also, as
demonstrated below, the regions where most of the conductive and
viscous dissipation take place do not have to be spatially
overlapping, i.e., some regions can be dominated by viscous or
conductive dissipation. \\

\indent The dissipation of mechanical energy injected by the AGN can
be estimated as follows.  Assuming that the temperature varies
smoothly with position in the fluid, the rate of change of mechanical
energy can be written as

\begin{eqnarray}
\int_{V}\dot{\epsilon}_{\rm mech}dV & = & \int_{V}T\frac{ds}{dt}dV = \int_{V}\dot{\epsilon}_{\rm visc}dV\nonumber\\ 
   & + &    \int_{S}\kappa {\bf \nabla}T\cdot d{\bf S} + \int_{V}\frac{\kappa}{T}(\nabla T)^{2}dV
\end{eqnarray}

\noindent
For random temperature fluctuations and $|\nabla T|\gg\Delta T/L$,
where $\Delta T$ is the mean temperature change over some typical
length $L$, the second term on the right hand side is small compared
to the last one if we take the spatial average of equation 7. In
particular, for temperature fluctuations, such as waves, that occur in
a constant background temperature, this term will vanish when averaged
over one wavelength.  In other words, if the temperature gradient is
dominated by temperature changes on small scales, we may locally
approximate the dissipation of mechanical energy by the last term in
equation 7. This is the approach that we adopted when comparing the
dissipation due to viscosity and conductivity with the radiative
cooling.

The simulations of the effects of viscosity and conductivity in three
dimensions require high spatial resolution. This can be achieved at
the expense of a relatively short simulation time (a few full AGN
activity cycles). We point out that the timestep imposed by the
transport coefficients on the simulation scales as $(\Delta
x)^{2}/(\kappa,\mu)$, where $\Delta x$ is the simulation
resolution. As this timestep scales more strongly with $\Delta x$ than
the standard hydrodynamical Courant condition, and because the
transport coefficients depend strongly on temperature, the constraints
on the timestep are more stringent that the ones obtained from the
standard condition.  This is why we chose to resort to comparing
instantaneous rates of heating instead of evolving the system for a
longer time at lower spatial resolution as the latter could prevent us
from capturing essential physics. However, we note that the actual
evolution of the gas in the simulations, including the energy transfer
and dissipation, does not use the above approximation.
We also point out that the actual heating of the cluster core is
greater than that estimated from this prescription. This is because
the spatially and temporally averaged temperature gradient in a
cooling flow cluster is positive within the cooling radius and the
heat transfer from the hot outer layers will take place. Thus, the
estimated conductive dissipation can be considered to be a lower limit
on the actual dissipation due to conductivity.

\section{Synthetic Chandra observations}

We have performed synthetic X-ray observations of the grid at two
stages of bubble evolution. These observations simulate a 200ks
ACIS-S3 exposure of the $(800 kpc)^3$ region centered on the bubble
origin. The images are generated by first calculating a MEKAL (Mewe et
al. 1985) emissivity at each grid point in three energy bands, then
integrating along each line-of-sight through the simulation box in the
optically thin limit using the SYNTH code (e.g., Tregillis et
al. 2004). The resulting X-ray surface brightness image in each band
is modified by galactic absorption, adjusted for the assumed redshift
of the simulated cluster (z=0.0183, Perseus) then processed through
the Chandra ACIS-S3 response function to generate instrument count
rates at each pixel of the image. The effect of vignetting is
included and corrected for in the final images. In this case the
image plane is larger than the dimensions of ACIS-S3, so we tile the
chip accordingly to cover the image plane. Since the angle
subtended by the smallest voxel in the simulation when placed at
Perseus' redshift is approximately 7.5\arcsec, no modeling of the PSF
is necessary.

To simulate the effect of the instrumental and X-ray background, we
have added the ACIS-S D period sky background to the resulting
images. The background event file is sorted to generate images in each
of the three X-ray bands, then binned to match image
resolution, and matched to the exposure of the synthetic
observations. To simulate the effect of background subtraction, we
have simply mirrored each of the three background images about the
vertical chip axis and subtracted the mirrored image in each case.
The net counts added and subtracted to the image then are the same,
but the counts in each pixel are not identical, resulting in the
appearance of a background-subtracted image. The final images have the
visual appearance of a Chandra X-ray image.  More importantly, in this
way we can quantify the expected error level associated with the
features in the images. \\
\indent
The final images are equivalent to exposure corrected, background
subtracted ACIS-S3 images of the simulated volume. In addition, for
the images in each of the three energy bands we have performed an
unsharp masking to generate an additional image. This image results
from subtracting an image smoothed with a 4 pixel gaussian kernel from
the original image. This procedure enhances the appearance of
fluctuations about the mean.

\section{Results}

\begin{figure*}
\begin{center}
\includegraphics[width=6.0in,angle=0]{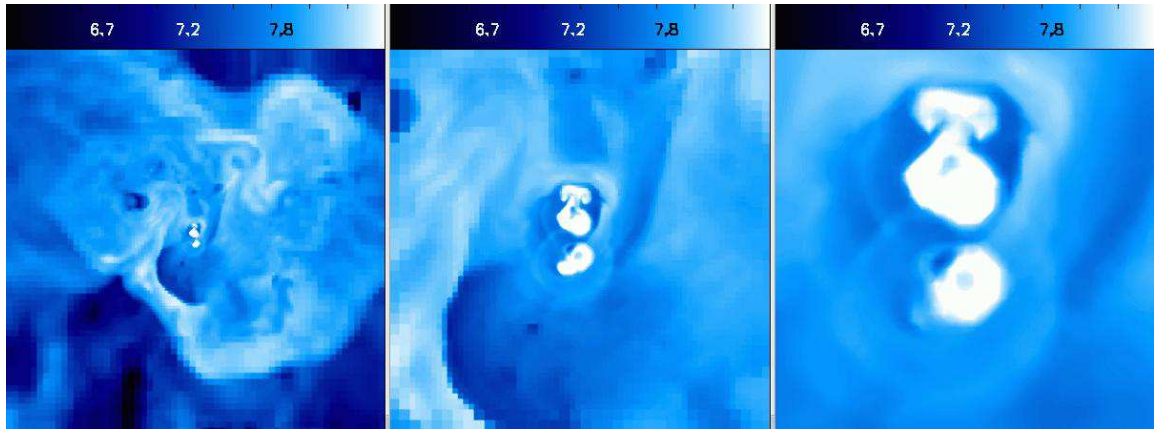}
\caption{Temperature distribution in a slice through the cluster
center. The box sizes in the panels are 2.8 Mpc, 800 kpc, and 300 kpc
from left to right, respectively. All panels correspond to the time of
140 Myr after the initial AGN outburst. Both young bubbles still in
the inflation phase as well as older buoyantly rising bubbles can be
seen. Note nearly spherical weak shocks surrounding the bubbles. All
plots are logarithmic and the colorbars show the logarithm of
temperature in K.}
\label{fig:temp}
\end{center}
\end{figure*}

Figure \ref{fig:temp} shows three slices through the cluster center displaying
the gas temperature at the time of 140 Myrs (file 0142) after the start of AGN
activity.  The panels have the size of 2.8 Mpc, 800 kpc and 300 kpc on a side
from left to right, respectively.  At this stage in the evolution one can
clearly identify these bubbles that have not yet reached pressure equilibrium
with their surroundings and that are still expanding nearly spherically into
the ambient medium (bubbles close to the injection region) as well as older
bubbles that have evolved into mushroom-type clouds.
As a result of the rapid inflation, the younger bubbles produce a weak shock
wave that can be seen to travel outward. Figure \ref{fig:initT} also shows
that the cluster is quite dynamic, as it shows significant substructure such
as clumps and shock fronts.

\begin{figure*}
\begin{center}
\includegraphics[width=6.0in,angle=0]{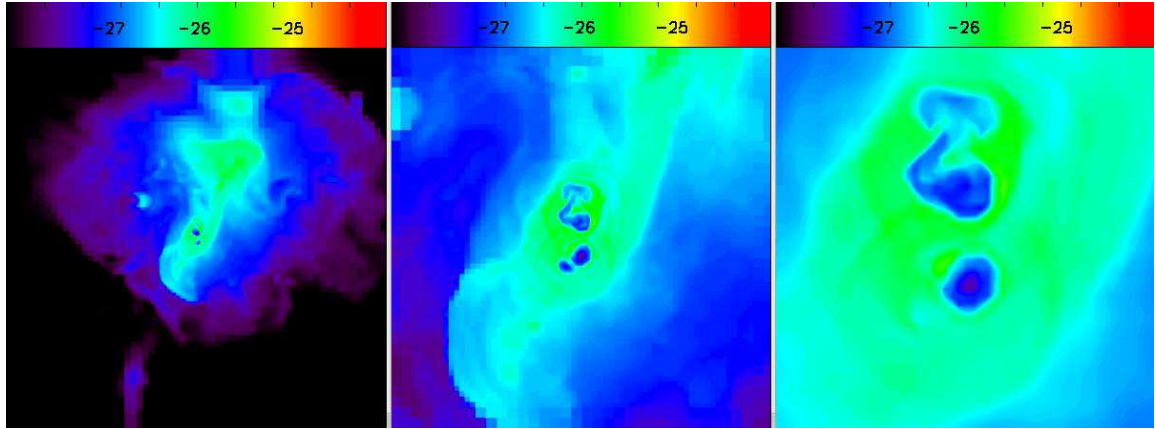}
\caption{Same as Fig.~\ref{fig:temp} but for physical gas density.}
\label{fig:dens}
\end{center}
\end{figure*}

Figure \ref{fig:dens} shows the density distribution in the
cluster. This figure corresponds to the same time and box sizes as
Figure \ref{fig:initT}.  As in Figure \ref{fig:initT} one can clearly
identify the young bubbles that are still in the expansion phase and
the older, buoyant bubbles.  As expected in the case of weak shocks,
the waves are weaker in the density maps than in temperature maps but
are still visible.\\ 

\indent Although the detached bubbles shown in Figures
\ref{fig:temp} and \ref{fig:dens} do not seem to be disrupted (their
detailed appearance depends on the exact position of the cross-section
plane), they eventually do tend to mix with the intracluster medium
late in the bubble evolution.  Observations indicate that the bubbles
maintain their identity as they move away from cluster centers. Most
observations detect bubbles in the early stages of their evolution
where various instabilities may not have had enough time to disrupt
them. However, if the real bubbles indeed preserve their identity at
larger distances then this effect is not explained by our
simulations. This discrepancy could then be explained if either the
bubbles themselves were viscous (as discussed by Reynolds et al. 2004)
or if they were magnetized in which case Rayleigh-Taylor and
Kelvin-Helmholtz instabilities on the bubble surfaces would be
suppressed. We note that even initially weak magnetic fields (i.e.,
dynamically unimportant ones) could be amplified around the bubbles
and effectively prevent instabilities from happening (Jones \& DeYoung
2005).

\begin{figure*}
\begin{center}
\includegraphics[width=6.0in,angle=0]{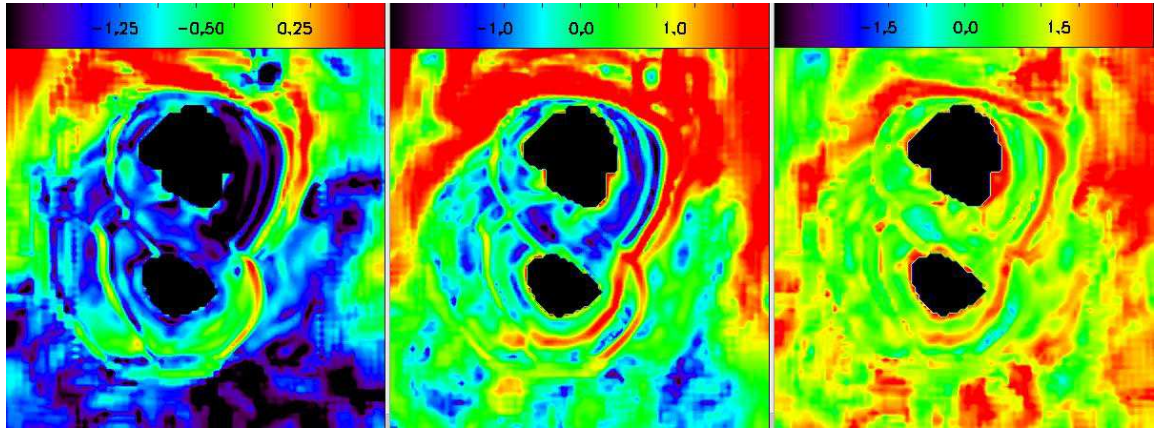}
\caption{Dissipation patterns. All panels are 300 kpc a side and
correspond to 80 Myr after the initial outburst of the AGN. Left panel
shows the logarithmic ratio of the viscous dissipation and the
radiative cooling rate. Middle panel presents the logarithm of the
ratio of the conductive dissipation to the radiative cooling rate and
the right one presents the logarithm of the ratio of the conductive
and viscous dissipation rates.}
\label{fig:diss1}
\end{center}
\end{figure*}

Figure \ref{fig:diss1} presents the dissipation rates corresponding to the time
of 80 Myr after the initial AGN outburst. The left panel shows the ratio of
the viscous dissipation to the radiative cooling.  As can clearly be seen, the
weak shocks present in the density and temperature maps are the sites of
enhanced dissipation. Some of the strongest waves have dissipation rates
comparable to the radiative cooling rates. The middle panel shows the ratio of
the conductive dissipation and the cooling rate. A comparison of this and the
left panel shows that the conductive dissipation in the shocks on average
appears to be higher. This is also visible in the right panel that shows the
ratio of the conductive and viscous dissipation rates and especially in the
profiles presented below.

\begin{figure*}
\begin{center}
\includegraphics[width=6.0in,angle=0]{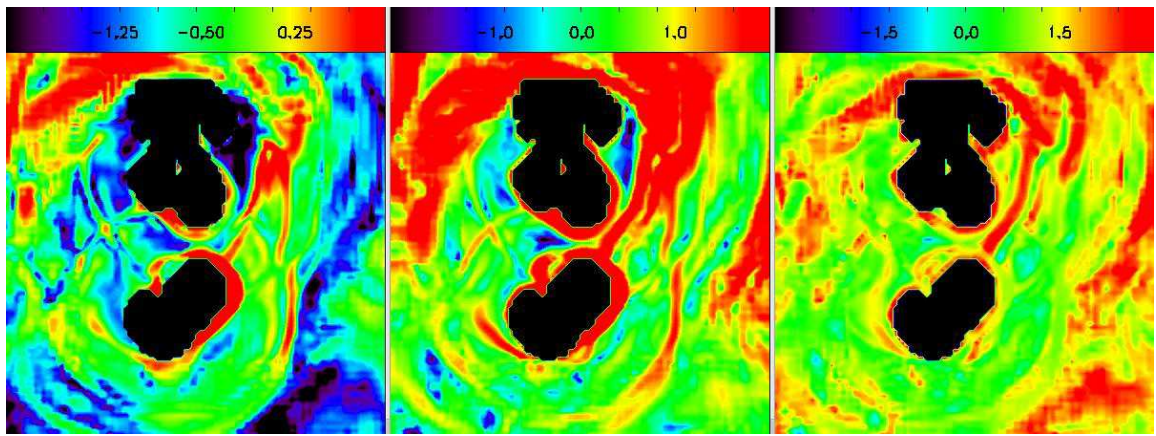}
\caption{Same as Figure \ref{fig:diss1} but for the simulation time of 140 Myr.}
\label{fig:diss2}
\end{center}
\end{figure*}

Figure \ref{fig:diss2} is analogous to Figure \ref{fig:diss1} but corresponds
to a later time (140 Myr) after the onset of AGN activity. The dissipation
patterns presented in this figure correspond to the same time as the
temperature and density maps in Figure \ref{fig:initT} and \ref{fig:dens},
respectively. Comparison of Figures \ref{fig:diss1} and \ref{fig:diss2}
shows that the dissipation patterns moved away from the center. The typical
speed of these patterns is of order of the sound speed in this cluster. This
reinforces the interpretation of these features as sound waves or weak shocks.

\begin{figure*}
\begin{center}
\includegraphics[width=6.0in,angle=0]{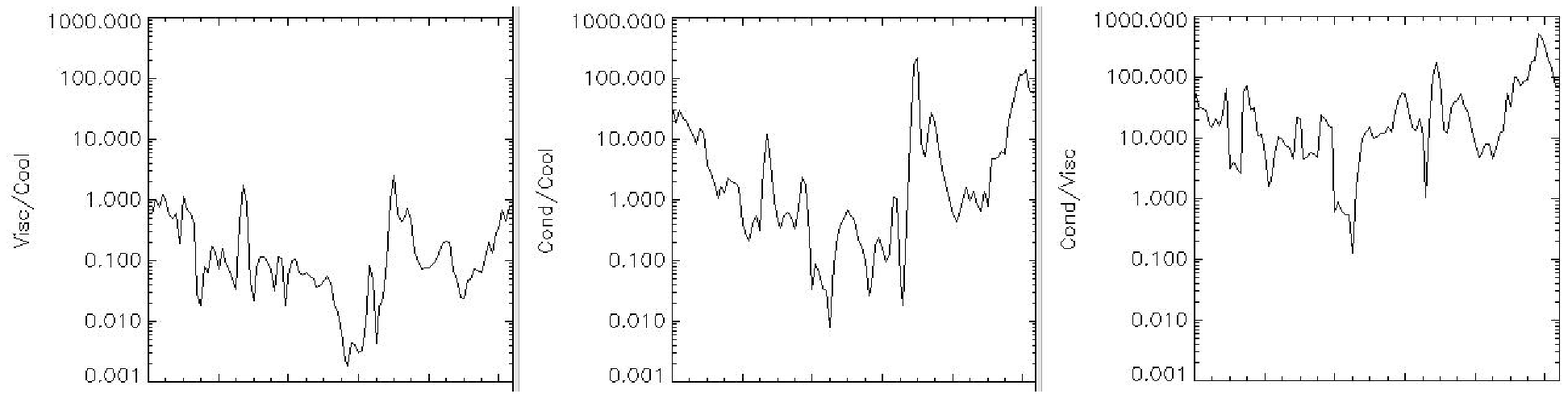}
\caption{Profiles along the horizontal lines through the center of the
dissipation patterns in Figure \ref{fig:diss1}.}
\label{fig:dissprof1}
\end{center}
\end{figure*}

\begin{figure*}
\begin{center}
\includegraphics[width=6.0in,angle=0]{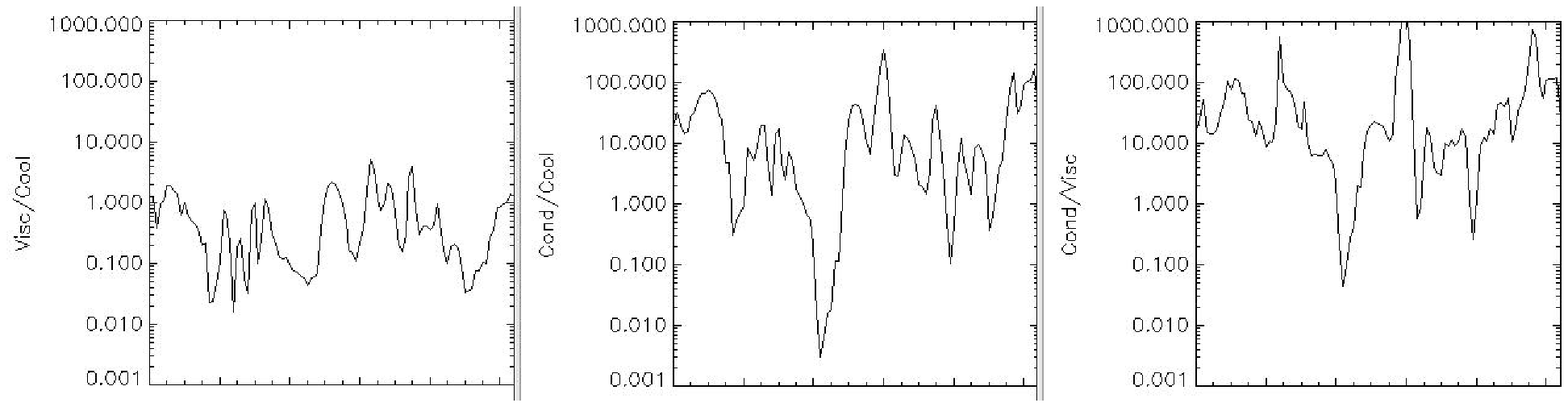}
\caption{Profiles along the horizontal lines through the center of the
dissipation patterns in Figure \ref{fig:diss2}.}
\label{fig:dissprof2}
\end{center}
\end{figure*}

Figures \ref{fig:dissprof1} and \ref{fig:dissprof2} present the profiles of
viscous and conductive dissipation for the time of 80 Myr and 140 Myr after
the first outburst, respectively. The profiles were extracted along the
horizontal lines intersecting the centers of the dissipation maps shown in
Figures \ref{fig:diss1} and \ref{fig:diss2}.  Left panels present the profiles
of the ratios of viscous dissipation and cooling rate.  The middle ones are
for conductive dissipation and the right panels show the profiles of the ratio
of conductive and viscous dissipation. It is interesting to note that the
conductive-to-viscous ratios seem to be $\sim 10$ which is consistent with
simple analytic argument presented in Section 3.3.\\
\indent
Figures \ref{fig:dissprof1} and \ref{fig:dissprof2} indicate that there is a large degree of scatter in the ratio of the heating rates. The deviations of the simulated ratios from the estimated constant ratio given by Equation 6 are mostly due to the fact that the density and temperature fluctuations are nonlinear (the gas may be effectively ``swept'' as a result of energy injection) and due to the the fact that bulk gas motions are present in the cluster. We note that the offsetting of the effects of cooling is helped by the fact that bubbles tend to transfer momentum to the ambient gas and disperse it. The gas that gets elevated and disperssed can be heated more efficiently. 

\begin{figure*}
\begin{center}
\includegraphics[width=6.0in,angle=0]{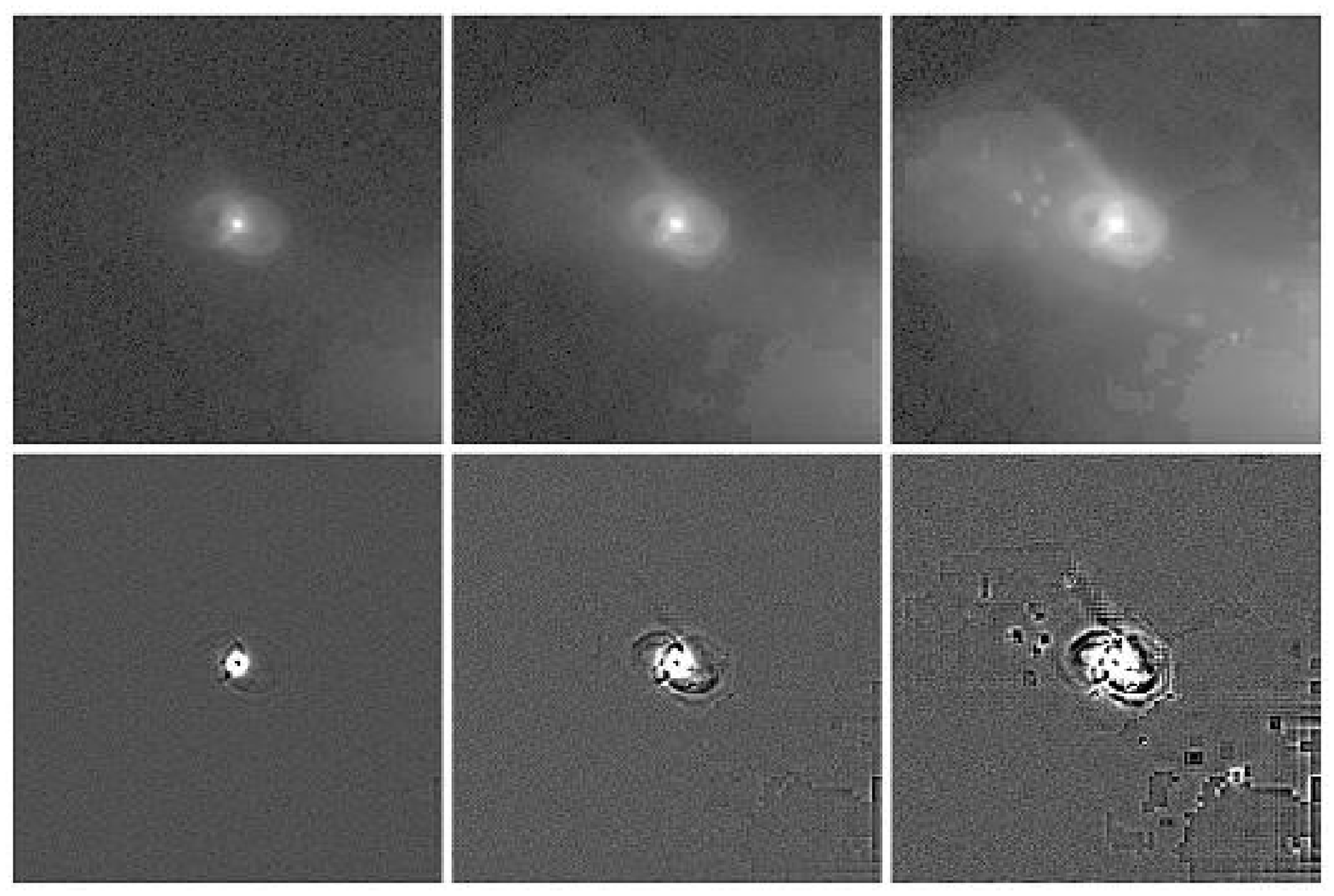}
\caption{X-ray images of the cluster heated by AGN at the time of 80
Myr. First and the second row show the maps of X-ray emissivity and
unsharped X-ray images, respectively. The columns are for [3.5-7.0]
keV [1.5-3.5] keV and [0.3-1.5] keV from left to right.}
\label{fig:xray1}
\end{center}
\end{figure*}

\begin{figure*}
\begin{center}
\includegraphics[width=6.0in,angle=0]{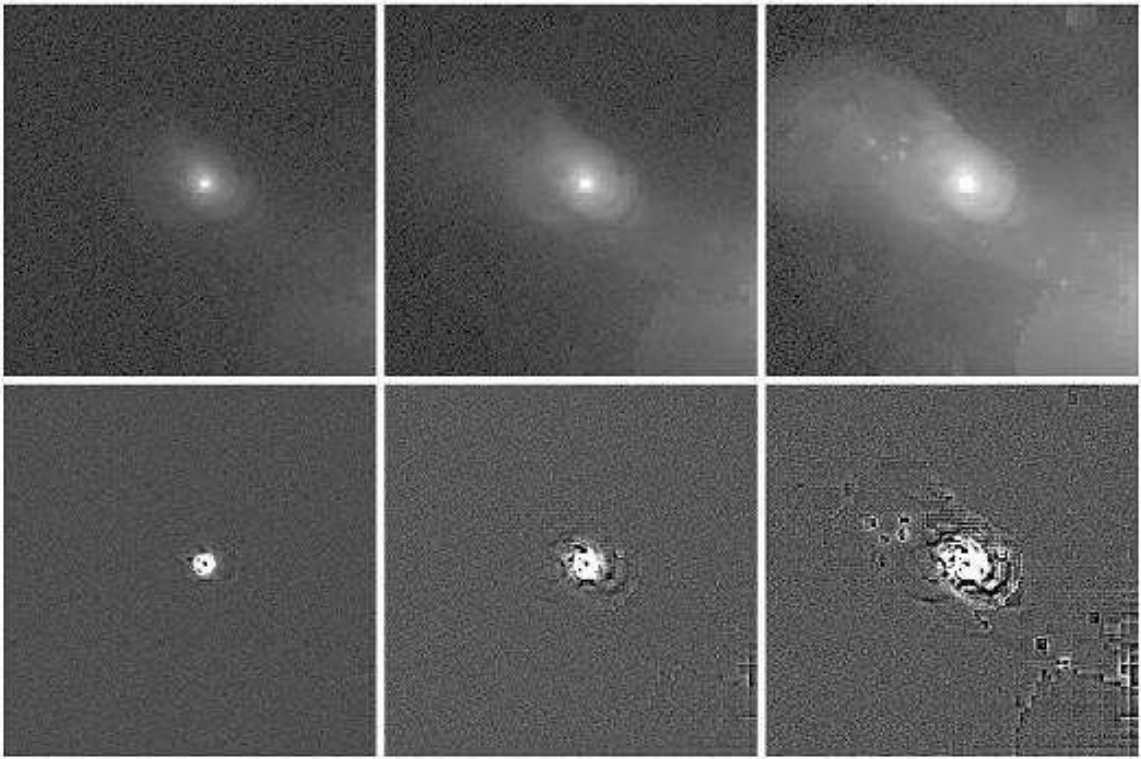}
\caption{Same as Figure \ref{fig:xray1} but for the time of 140 Myr.}
\label{fig:xray2}
\end{center}
\end{figure*}

Figures \ref{fig:xray1} and \ref{fig:xray2} show X-ray images of the cluster
heated by AGN at the time of 80 Myr and 140 Myr, respectively.  The first and
the second row show the maps of X-ray surface brighness and unsharped X-ray
images, respectively. The columns are for [3.5-7.0] keV [1.5-3.5] keV and
[0.3-1.5] keV from left to right.  These figures show clear evidence for X-ray
featureas at the spatial locations of the waves resulting from expansion of
the bubbles. The bubbles and waves can be clearly seen in the earlier stages
in the AGN evolution.  All features are more easily detectable in the softer
X-ray bands.

\begin{figure*}
\begin{center}
\includegraphics[width=6.0in,angle=0]{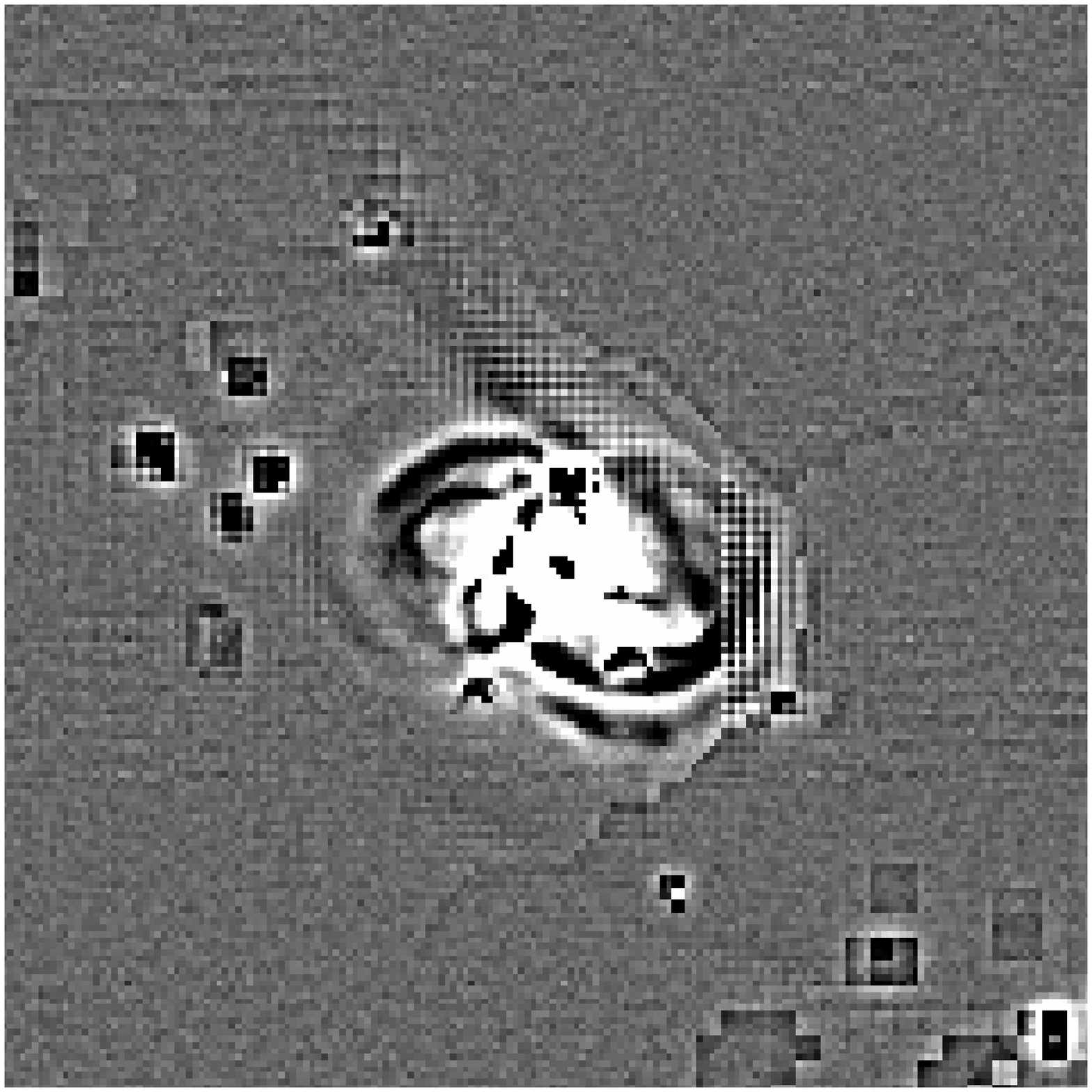}
\caption{Closeup of the bottom right panel in
Fig.~\ref{fig:xray1}. Both the waves and the ``bullet-like'' feature
are visible.}
\label{fig:closeup}
\end{center}
\end{figure*}

Figure \ref{fig:closeup} shows a closeup of the unsharp masked image in the
soft band corresponding to the time of 80 Myr (bottom right panel in Figure
\ref{fig:xray1}). Apart from the waves one can easily identify a bullet-like feature
wrapping around one side of the maximum in density distribution suggesting a
motion to the right.

\section{Summary}

We have simulated the effects of heating by active galactic nuclei on the
intracluster medium including, both, the effect of viscosity and conductivity.
This has been done by a high resolution 3D AMR simulation of a galaxy cluster
that has been extracted from full a cosmological SPH simulation.  Our
simulations demonstrate that conductivity is likely to play an important role
in dissipating mechanical energy injected by the AGN.  We have shown that if
the suppression of conductivity and viscosity is of comparable magnitude, then
the effect of conductivity on dissipation is likely to exceed that due to
viscosity. Nevertheless, viscosity may play a significant role in affecting
the dynamical evolution of the buoyant bubbles especially if suppression of
conductivity is higher than that of viscosity.  The precise values of
suppression factors are unknown as yet but we hope that future observations of
the ICM would be able to put constraints on their values. 
Our simulations were performed for a moderate mass cluster. As higher mass clusters have higher temperatures and densities and the opposite is true for lower mass ones, we expect that qualitatively the trends found in our simulations could hold for clusters different than the one considered here. However, we note that the details of feedback are sensitive to the amount of the injected energy that may vary depending on the cluster mass (Roychowdhury et al. 2004). They also depend sensitively on the exact values of density and temperature as they enter in high powers in the expressions for cooling and dissipation rates.
We also discuss the
detectability of the X-ray features generated by AGN outbursts. Our synthetic
data takes into account all instrumental features of the {\it Chandra}
Observatory and all sources of noise such as X-ray background.  We show that
bubbles inflated by AGN and the waves generated by the expansion of the
bubbles can be seen in these synthetic observations.  The detactability of
these features depends on the stage in the AGN evolution (with bubbles more
easily detectable in the earlier stages) and the {\it Chandra} energy band
observed (with stronger features in softer bands).  Other features such as 
bullets due to substructure motions are also detectable in the simulated X-ray
emissivity maps. These are similar to that observed in 1E 0657-56 (Markevitch
et al. 2002)

\section{Acknowledgments}
We thank Mitch Begelman, Jim Pringle and Mike Norman for very helpful discussions on this topic and
Volker Springel for providing us with some of his simulations.
We would also like to thank an
anonymous referee for constructive comments on the text.
Support for this work was provided by National Science Foundation grant
AST-0307502 and NASA through {\it Chandra} Fellowship Award Number PF3-40029
issued by the Chandra X-ray Observatory Center, which is operated by
the Smithsonian Astrophysical Observatory for and on behalf of NASA 
under contract NAS8-39073.
MB gratefully acknowledges support by DFG grant BR 2026/2 and the
supercomputing grant NIC 1658. The software used in this work was in
part developed by the DOE-supported ASCI/Alliance Center for
Astrophysical Thermonuclear Flashes at the University of Chicago.
Calculations presented in this paper were in part performed at National Center
for Supercomputing Applications at the University of Illinois at 
Urbana-Champaign, which is founded by the PACI Program at the National
Science Foundation (Alliance Allocations Board grant number AST040015).
The help from the NCSA technical support team is greatly acknowledged.
Eric Hallman acknowledges the support from 
NSF grant AST-0407368.

\end{document}